  \documentclass[apj,amsmath,amssymb]{emulateapj-rtx4}
\usepackage{color}
\begin{document}


\title{Radiative transfer simulation for the optical and near-infrared electromagnetic counterparts to GW170817}


\author{Kyohei Kawaguchi}
\affil{Institute for Cosmic Ray Research, The University of Tokyo, 5-1-5 Kashiwanoha, Kashiwa, Chiba 277-8582, Japan}
\affil{Max Planck Institute for Gravitational Physics (Albert Einstein Institute), Am M\"{u}hlenberg 1, Potsdam-Golm, 14476, Germany}\affiliation{Center for Gravitational Physics,
  Yukawa Institute for Theoretical Physics, 
Kyoto University, Kyoto, 606-8502, Japan} 
\author{Masaru Shibata}
\affiliation{Center for Gravitational Physics,
  Yukawa Institute for Theoretical Physics, 
Kyoto University, Kyoto, 606-8502, Japan} 
\affil{Max Planck Institute for Gravitational Physics (Albert Einstein Institute), Am M\"{u}hlenberg 1, Potsdam-Golm, 14476, Germany}
\and
\author{Masaomi Tanaka}
\affil{Astronomical Institute, Tohoku University, Aoba, Sendai 980-8578, Japan}
\affil{National Astronomical Observatory of Japan, Mitaka, Tokyo, Japan}


\newcommand{\angstrom}{\text{\normalfont\AA}}
\newcommand{\rednote}[1]{{\color{red} (#1)}}

\begin{abstract}
Recent detection of gravitational waves from a binary-neutron star merger (GW170817) and the subsequent observations of electromagnetic counterparts provide a great opportunity to study the physics of compact binary mergers. The optical and near-infrared counterparts to GW170817 (SSS17a, also known as AT 2017gfo or DLT17ck) are found to be consistent with a kilonova/macronova scenario with red and blue components. However, in most of previous studies in which contribution from each ejecta component to the lightcurves is separately calculated and composited, the red component is too massive as dynamical ejecta and 
the blue component is too fast as post-merger ejecta. In this letter, we perform a 2-dimensional radiative transfer simulation for a kilonova/macronova consistently taking the interplay of multiple ejecta components into account. We show that the lightcurves and photospheric velocity of SSS17a can be reproduced naturally by a setup consistent with the prediction of the numerical-relativity simulations. %Furthermore, we find that the mass-averaged element abundance of the ejecta model agrees broadly with the solar abundance of r-process element. 
\end{abstract}



\section{Introduction}
On 17th August 2017, the first detection of gravitational waves from a binary neutron star (NS) merger referred to as GW170817 was achieved by three ground-based detectors~\citep{TheLIGOScientific:2017qsa}. Electromagnetic (EM) counterparts to GW170817 were observed over the entire wavelength range. Gamma-ray signals were detected about 1.7 seconds after the onset of the merger~\citep{Monitor:2017mdv}, and then, a counterpart in ultraviolet, optical, and near-infrared (NIR) wavelengths, named as SSS17a (also known as AT 2017gfo or DLT17ck), is discovered ~\cite[e.g.,][see more references in ~\cite{Villar:2017wcc}]{Coulter:2017wya,Valenti:2017ngx,Drout:2017ijr,Cowperthwaite:2017dyu,Kasliwal:2017ngb,Smartt:2017fuw,Tanvir:2017pws} . NGC 4993, a galaxy at a distance of 40 Mpc, was identified as the host galaxy of GW170817 by the EM signals. X-ray~\citep[e.g.,][]{Troja:2017nqp} and radio signals~\citep[e.g.,][]{Mooley:2017enz} were also detected subsequently.

Among various EM signals from NS mergers, the emission in optical and NIR wavelengths is in particular of interest. It has been suggested that a fraction of NS material would be ejected from the system during the merger~\cite[e.g.,][]{Rosswog:1998hy,Hotokezaka:2012ze}, and heavy radioactive nuclei would be synthesized in the ejecta by the so-called {\it r-process} nucleosynthesis~\citep{Lattimer:1974slx,Eichler:1989ve,Korobkin:2012uy,Wanajo:2014zka}. It has been predicted that EM emission in optical and NIR wavelengths could occur by radioactive decays of heavy elements~\citep{Li:1998bw,Kulkarni:2005jw,Metzger:2010sy,Kasen:2013xka,Tanaka:2013ana}. This emission is called ``kilonova'' or ``macronova''. Previous studies~\citep{Li:1998bw,Kasen:2013xka,Kasen:2014toa,Barnes:2016umi,Wollaeger:2017ahm,Tanaka:2017qxj,Tanaka:2017lxb} showed that lightcurves of kilonovae/macronovae depend on the mass, velocity, and electron fraction ($Y_e$, number of protons per nucleon which controls the final element abundances) of ejecta.  These quantities reflect the mass ejection mechanism, and thus, we can study the physical process of NS merger and associated r-process nucleosynthesis via detailed analysis of kilonovae/macronovae lightcurves.

Several ejection mechanisms are proposed for NS mergers. One is called the dynamical ejection, which is driven by tidal interaction and shock heating during the collision of NSs~\citep{Hotokezaka:2012ze,Bauswein:2013yna,Sekiguchi:2016bjd,Radice:2016dwd,Dietrich:2016hky,Bovard:2017mvn}. Numerical relativity simulations for binary NS mergers show that the mass and {\em averaged} velocity of the dynamical ejecta are typically $10^{-3}$--$10^{-2}\,M_\odot$ and $0.15$--$0.25\, c$, respectively, where $c$ is the speed of light. The electron fraction is distributed in the range of $0.05$--$0.5$, which leads to a large value of opacity $\sim 10\,{\rm cm^2/g}$~\citep{Kasen:2013xka,Tanaka:2013ana}. Due to such high opacity, the kilonova/macronova emission from the dynamical ejecta is expected to be bright in NIR wavelengths and last for $\sim10$ days (hereafter we refer to it as the red component).  After the dynamical ejection, the mass ejection from the merger remnant driven by viscous and neutrino heating can follow~\citep[][we refer to these ejecta as post-merger ejecta.]{Dessart:2008zd,Metzger:2014ila,Just:2014fka,Siegel:2017nub,Shibata:2017xdx,Fujibayashi:2017puw} Numerical-relativity simulations considering the effects of physical viscosity and neutrino radiation show that $10^{-2}$--$10^{-1}\,M_\odot$ of the material can be ejected typically with the velocity of $\alt0.1\,c$ from the massive NS and torus formed after the merger. Due to the irradiation by neutrinos emitted from the remnant NS, the electron fraction of the post-merger ejecta typically has a larger value ($Y_e\approx0.3$--$0.4$) than that of the dynamical ejecta~\citep{Metzger:2014ila,Fujibayashi:2017puw}. This leads to a smaller value of opacity $\sim0.1$--$1\,{\rm cm^2/g}$~\citep{Kasen:2014toa,Tanaka:2017lxb}, and hence, blue optical emission which lasts for $\sim1$ day would occur (hereafter we refer to it as the blue component).

 A number of studies have shown that SSS17a is consistent with kilonova/macronova models composed of red and blue (or more) components~\citep[e.g.,][]{Kasliwal:2017ngb,Cowperthwaite:2017dyu,Kasen:2017sxr,Villar:2017wcc}.  However, (i) the estimated mass for the red component, $10^{-2}$--$10^{-1}\,M_\odot$, is more massive than the theoretical prediction for the dynamical ejecta $(\alt 0.01 M_\odot)$~\citep{Hotokezaka:2012ze,Bauswein:2013yna,Sekiguchi:2016bjd,Radice:2016dwd,Dietrich:2016hky,Bovard:2017mvn}, and (ii) ejecta velocity $\agt 0.1$--$0.3\,c$ required for the blue component is too high for the post-merger ejecta found in numerical-relativity simulations~\citep[e.g.,][which show typically $\sim0.05\,c$]{Metzger:2014ila,Fujibayashi:2017puw}. 

In these kilonovae/macronovae models~\citep{Kasliwal:2017ngb,Cowperthwaite:2017dyu,Kasen:2017sxr,Villar:2017wcc}, contribution from each ejecta component to the lightcurves is separately calculated and composited. However, in reality, the lightcurves are determined through the non-trivial radiation transfer of photons in both ejecta components. In this letter, we perform an axisymmetric radiative transfer simulation for kilonovae/macronovae taking the interplay of multiple ejecta components of non-spherical morphology into account. We show that the optical and NIR lightcurves of SSS17a can be reproduced by the ejecta model which agrees quantitatively with the prediction of numerical-relativity simulations.

\section{Method and Model}
We derive lightcurves and spectra of kilonovae/marconovae by a wavelength-dependent radiative transfer simulation~\citep{Tanaka:2013ana,Tanaka:2017qxj,Tanaka:2017lxb}. The photon transfer is calculated by the Monte Carlo method for given ejecta profiles of density, velocity, and element abundance. The nuclear heating rates are given based on the results of r-process nucleosynthesis calculations by~\cite{Wanajo:2014zka}. We also consider the time-dependent thermalization efficiency following an analytic formula derived by~\cite{Barnes:2016umi}. We update the code so that special-relativistic effects on photon transfer are fully taken into account. The grid resolution of the simulation is also improved by an oder of magnitude from our previous works by imposing axisymmetry. 

For photon-matter interaction, we consider the same physical processes as in~\cite{Tanaka:2013ana,Tanaka:2017qxj,Tanaka:2017lxb}. Bound-bound, bound-free, and free-free transitions and electron scattering are considered for a transfer of optical and NIR photons. For the bound-bound transitions, which have a dominant contribution in the optical and NIR wavelengths, we used the formalism of the expansion opacity~\citep{1993ApJ...412..731E,Kasen:2006ce}.  For atomic data, the same line list as in~\cite{Tanaka:2017qxj} is used. This line list is constructed by the atomic structure calculations for Se ($Z = 34$), Ru ($Z = 44$), Te ($Z = 52$), Nd ($Z = 60$), and Er ($Z = 68$) and supplemented by Kurucz's line list for $Z < 32$~\citep{1995all..book.....K}. Since the atomic data are not complete,
we assume the same bound-bound transition properties for the elements with the same open shell as in \cite{Tanaka:2017lxb}. Since the atomic data include only up to doubly ionized ions, our calculations are applicable only for $\agt0.5$ days after the merger, for which the temperature is low enough ($\alt10000\,{\rm K}$). The ionization and excitation states are calculated under the assumption of local thermodynamic equilibrium by using the Saha ionization and Boltzmann excitation equations.

Numerical-relativity simulations give a picture that the post-merger ejecta is surrounded by the dynamical ejecta because the latter has higher velocity than the former. For such a situation, the post-merger ejecta would irradiate and heat up the dynamical ejecta, and help the long-lasting NIR lightcurves to be reproduced by less massive dynamical ejecta. Furthermore, since the dynamical ejecta has higher velocity than the post-merger ejecta, the reprocess of photons in the dynamical ejecta helps the photospheric velocity to be enhanced. Most of the dynamical ejecta is present near the binary orbital plane (i.e., $\theta \agt \pi/4$), and only a part of the dynamical ejecta is present in the polar region ($\theta\le\pi/4$), where $\theta$ is the inclination angle measured from the orbital axis of the binary~\citep[e.g.,][]{Hotokezaka:2012ze,Sekiguchi:2016bjd,Radice:2016dwd}. Nevertheless, low-density dynamical ejecta in the polar region can significantly modify the spectrum due to large opacity determined by lanthinides as known as the lanthanide curtain effect~\citep{Kasen:2014toa,Wollaeger:2017ahm}. Since the gravitational-wave data analysis of GW170817 infers that the event was observed from $\theta\alt28^\circ$~\citep{TheLIGOScientific:2017qsa}, photon-reprocessing in both the low-density and high-density dynamical ejecta would be important for the lightcurve prediction.

\begin{figure}
 	 \includegraphics[width=1\linewidth]{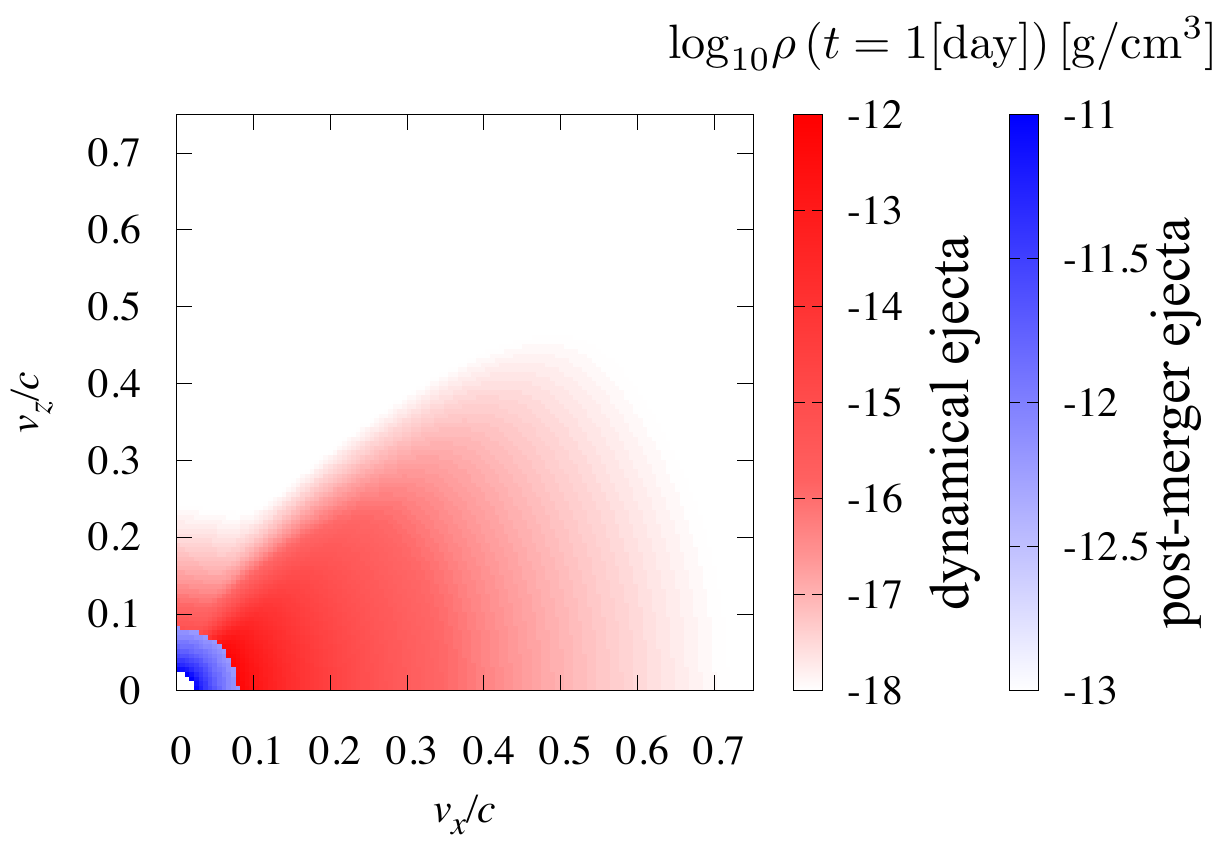}
 	 \caption{Density profile of the ejecta employed in the radiative transfer simulation. The red and blue regions denote the dynamical and post-merger ejecta, respectively. Homologous expansion of the ejecta and axisymmetry with respect to the $z$-axis are assumed in the simulation.}\label{fig:picture}
\end{figure}

For our radiative transfer simulation, we need the density and velocity profiles of ejecta. Within the range of predictions by numerical-relativity simulations, we set up a model which reproduces the key observational data of SSS17a, such as the lightcurves and photospheric velocity. We employ homologously expanding ejecta composed of two parts; the post-merger ejecta with the velocity from $v=0.025\,c$ to $0.08\,c$ and the dynamical ejecta from $v=0.08\,c$ to $0.9\,c$, where $v=r/t$ is the velocity of the fluid elements, $r$ is the radius, and $t$ is time measured from the onset of the merger. Note that the presence of the high-velocity components with $v \agt 0.3\,c$ up to $\sim 0.9\,c$ is found and confirmed by the latest high-resolution numerical-relativity simulation~\citep{Hotokezaka:2018gmo}. We adopt a power-law density distribution of $\propto r^{-3}$ and $\propto r^{-6}$ for the post-merger and dynamical ejecta, respectively, following the numerical-relativity results. To take the morphology of the dynamical ejecta into account, the density for $\theta\le\pi/4$ is set to be $\approx1000$ times smaller than that for $\theta\ge\pi/4$, and the low- and high-density regions are smoothly connected employing a logistic function, $\left\{1+{\rm exp}\left[-20\left(\theta-\pi/4\right)\right]\right\}^{-1}$ (see Figure~\ref{fig:picture}.) The total masses of the post-merger and dynamical ejecta are set to be $0.02\,M_\odot$ and $\approx0.009\,M_\odot$, respectively. The latest numerical-relativity simulations show that these are reasonable values~\citep[e.g.,][]{Hotokezaka:2012ze,Dietrich:2016hky,Metzger:2014ila,Fujibayashi:2017puw}. Following the numerical-relativity results~\citep{Sekiguchi:2016bjd,Shibata:2017xdx,Fujibayashi:2017puw}, the element abundances are determined by r-process nucleosynthesis calculations by~\cite{Wanajo:2014zka} assuming flat $Y_e$ distributions from $0.3$--$0.4$ and $0.1$--$0.4$ for the post-merger and dynamical ejecta, respectively. Note that the post-merger ejecta often has a component of $Y_e \agt 0.4$, which does not contribute significantly to heating because the heavy elements are not synthesized from such component~\citep{Wanajo:2014zka,Kasen:2014toa}. Here, the mass of $0.02M_\odot$ required for the post-merger ejecta is for the component with $Y_e \alt 0.4$. 

\section{Results}
\begin{figure*}
 	 \includegraphics[width=.475\linewidth]{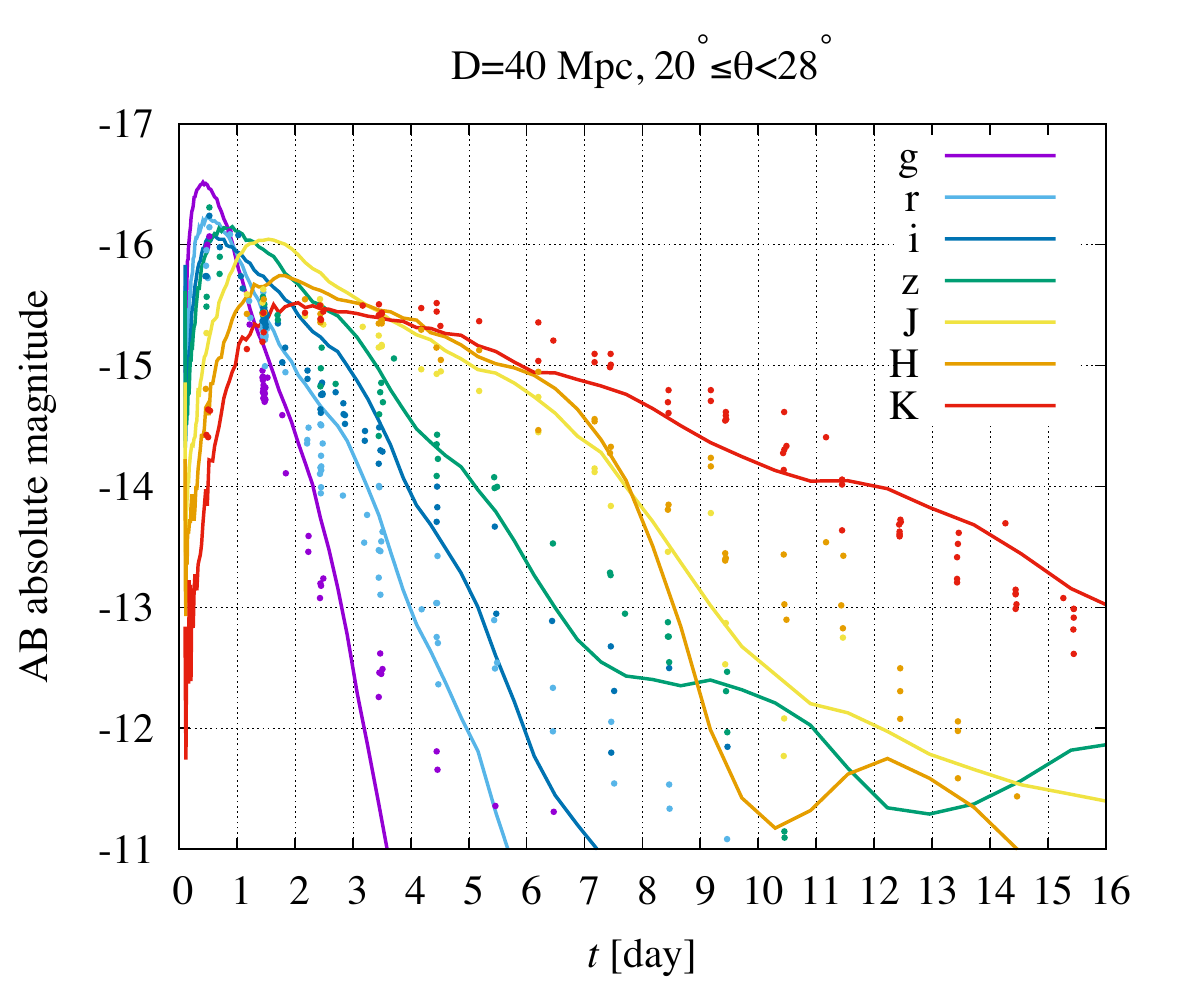}
 	 \includegraphics[width=.475\linewidth]{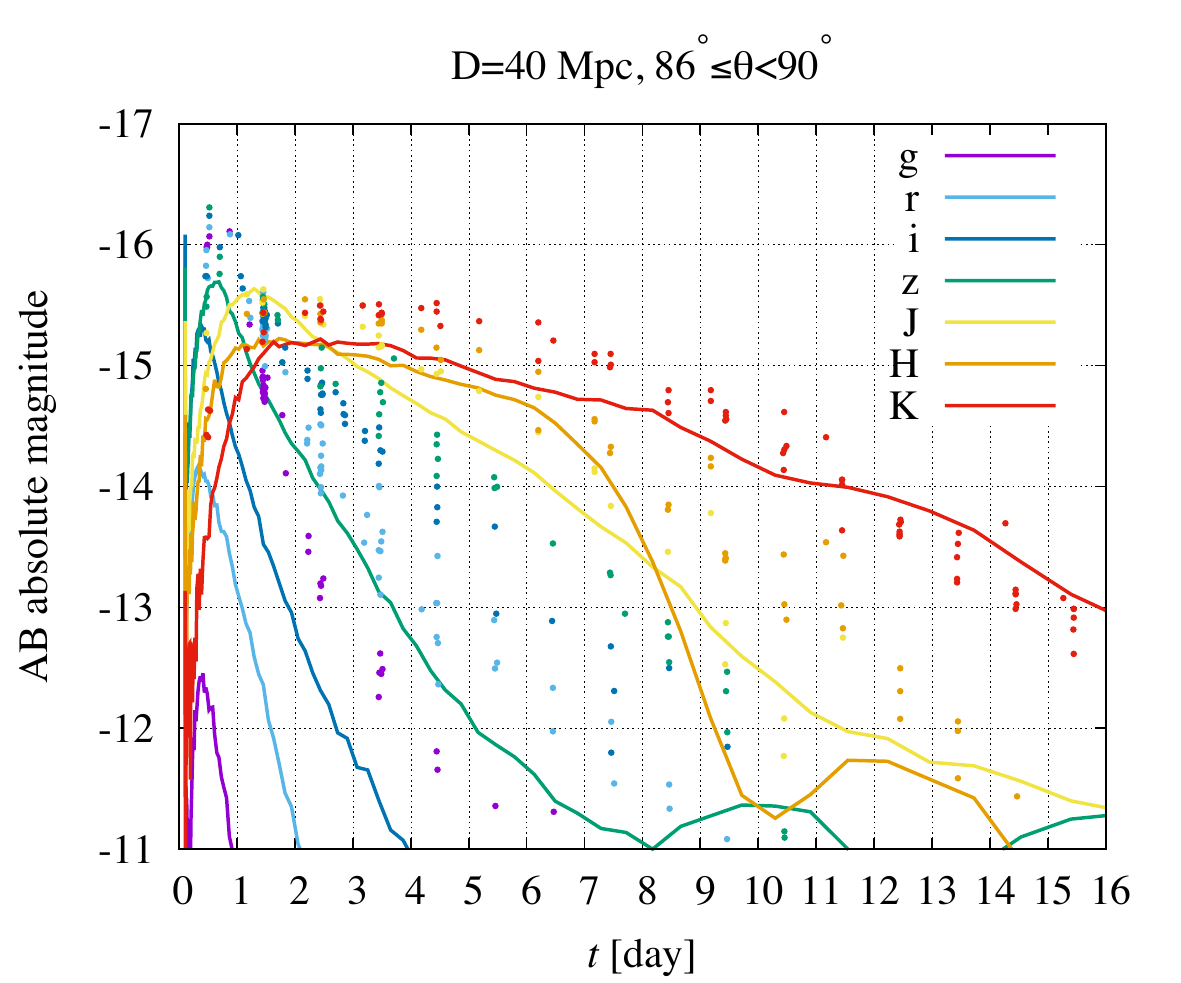}
 	 \caption{Optical and NIR lightcurves of SSS17a compared with the kilonova model observed from $20^\circ\le\theta\le28^\circ$ (left panel) and $86^\circ\le\theta\le90^\circ$ (right panel). The optical and NIR data points are taken from~\cite{Villar:2017wcc}. We assume that SSSa17 is at a distance of 40 Mpc. All the magnitudes are given in AB magnitudes. Note that the large deviation of the model lightcurves in {\it H}-band may be due to the incompleteness of the line list for the opacity estimation.}\label{fig:mag}
\end{figure*}

Figure~\ref{fig:mag} compares the observed {\it ugrizJHK}-band lightcurves of SSS17a~\citep{Villar:2017wcc} and those of our kilonova/macronova model. As a fiducial model to interpret the lightcurves of SSS17a, we employ the lightcurves observed from $20^\circ\le\theta\le28^\circ$ taking into account the results of the gravitational-wave data-analysis of GW170817~\citep{TheLIGOScientific:2017qsa}. We find that both optical and NIR lightcurves of SSS17a are approximately reproduced by a setup motivated by numerical-relativity simulations. In particular, the {\it ugri} and {\it zJHK}-band lightcurves of the model agree with the data points within 1 mag for $t\le2.5$ days and $t\le9$ days, respectively.   

In our model, the long-lasting NIR lightcurves are reproduced by the dynamical ejecta of which mass is much smaller than that estimated by the previous studies employing a simple composited model of ejecta components~\citep[e.g.,][]{Kasliwal:2017ngb,Cowperthwaite:2017dyu,Kasen:2017sxr,Villar:2017wcc}. This can be understood by the irradiation from the post-merger ejecta to the dynamical ejecta. The mass of the post-merger ejecta is also smaller than that estimated by previous studies. This is due to the diffusion of photons preferentially to the polar direction by which the luminosity is effectively enhanced in the polar direction in the presence of the optically thick dynamical ejecta in the equatorial plane. Indeed, we find that the total luminosity integrated over all the viewing angles is smaller by a factor of 2--3 than the isotropic luminosity observed from $20^\circ\le\theta\le28^\circ$~\citep[see also][]{Kasen:2014toa}. 

The right panel of Figure~\ref{fig:mag} shows the lightcurves of the model observed from the equatorial direction ($86^\circ\le\theta\le90^\circ$). The {\it ugriz}-band luminosity is much smaller than that observed from $20^\circ\le\theta\le28^\circ$, while similar magnitudes of luminosity are found in the {\it JHK}-bands. This reflects the fact that photons from the post-merger ejecta are entirely absorbed by the dynamical ejecta concentrated in the equatorial plane. This suggests that bright emissions in {\it ugriz}-band as found in SSS17a would not be observed for a similar NS merger if it is observed from the direction of the orbital plane.

\begin{figure}
 	 \includegraphics[width=1\linewidth]{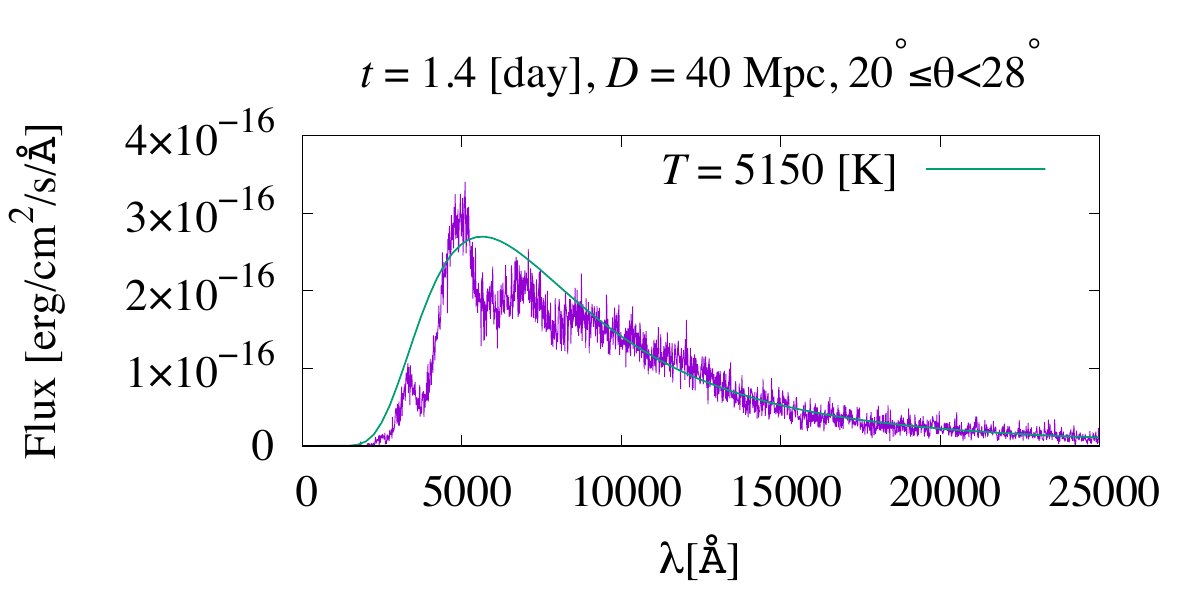}\\
 	 \includegraphics[width=1\linewidth]{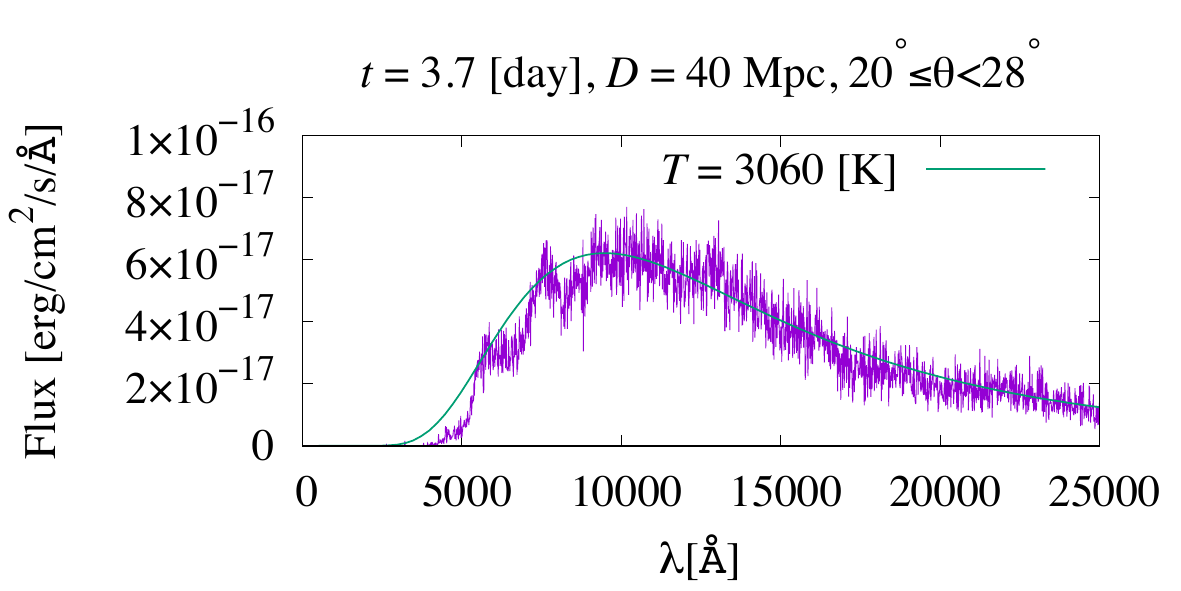}\\
 	 \includegraphics[width=1\linewidth]{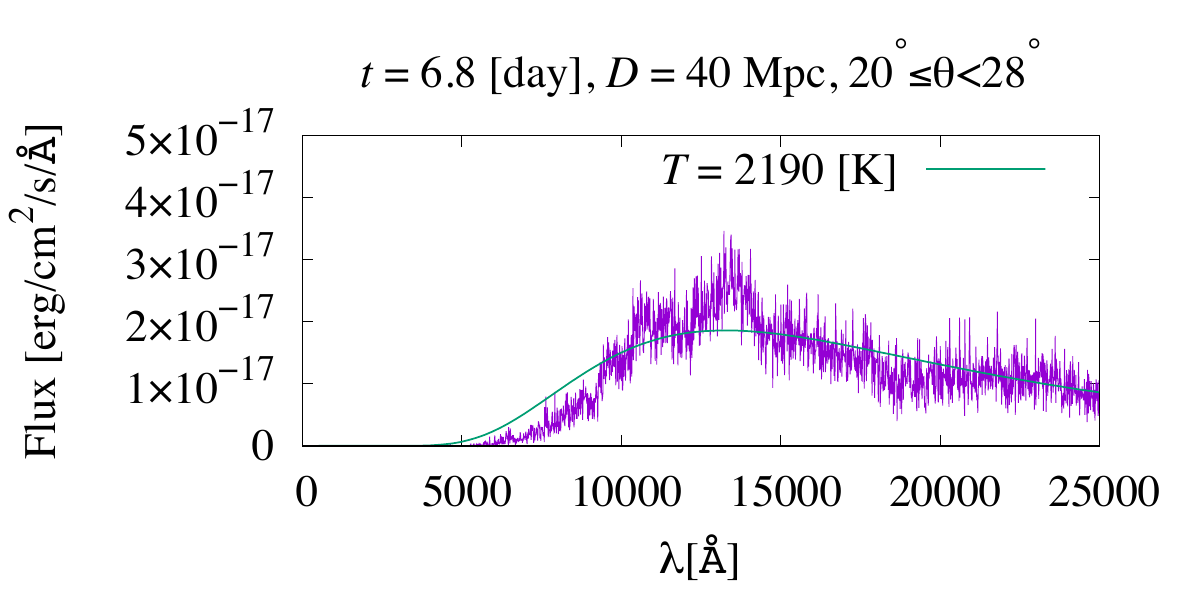}
 	 \caption{Time evolution of optical and NIR spectral energy distribution of the kilonova/macronova model. The spectra at $t=$1.4, 3.7, and 6.8 days are shown. All the spectra are observed from $20^\circ\le\theta\le28^\circ$ at a distance of 40 Mpc. The green solid curves denote the best blackbody fits of the spectra.}\label{fig:spec}
\end{figure}

\begin{figure*}
 	 \includegraphics[width=.45\linewidth]{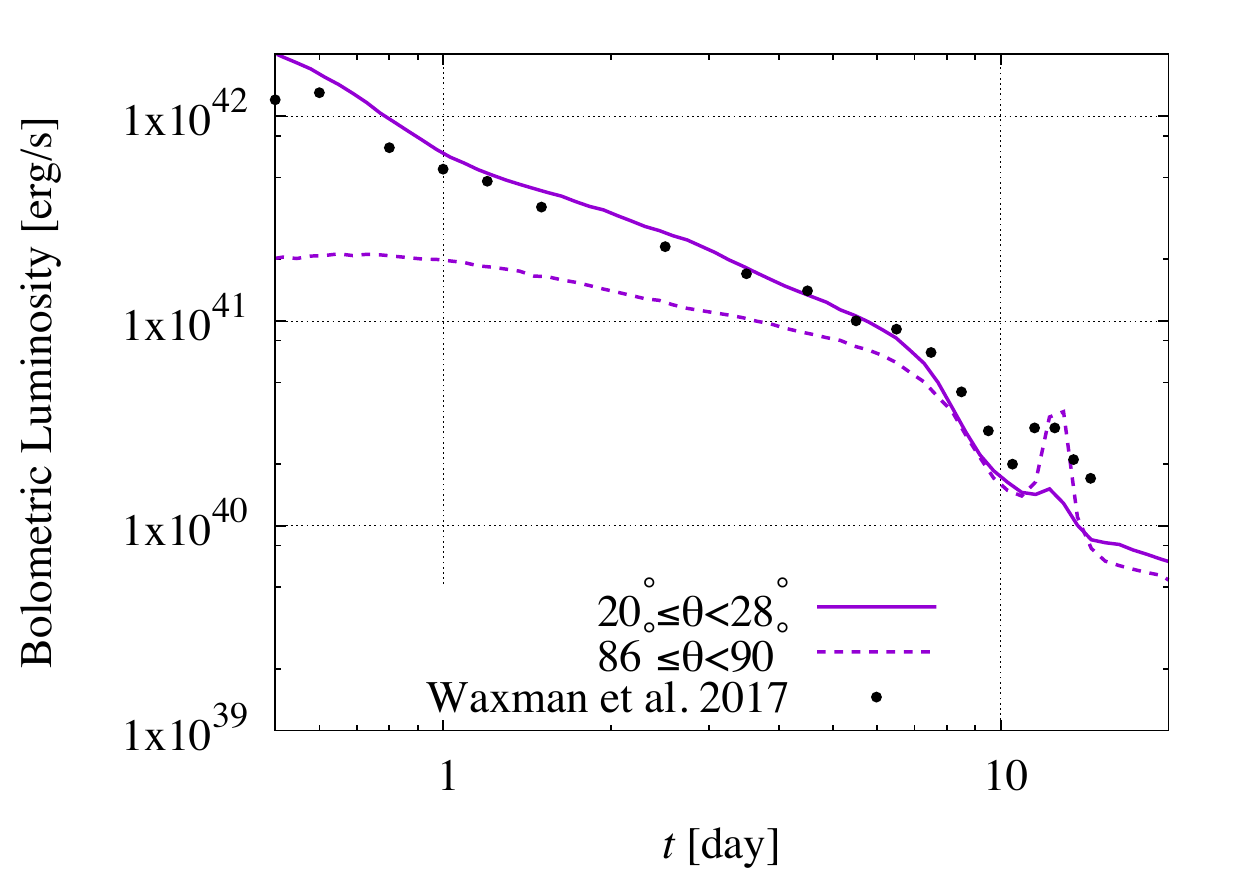}
 	 \includegraphics[width=.45\linewidth]{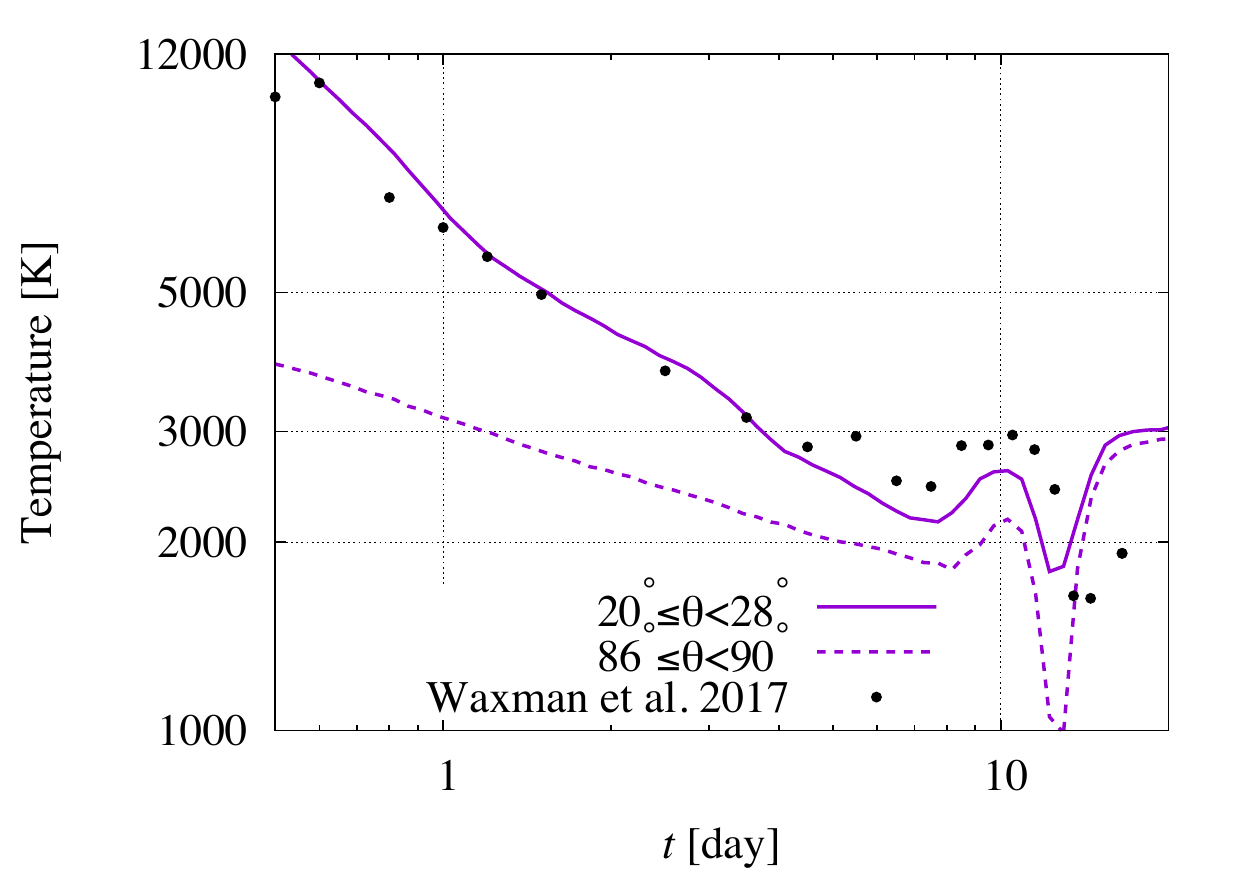}\\
 	 \includegraphics[width=.45\linewidth]{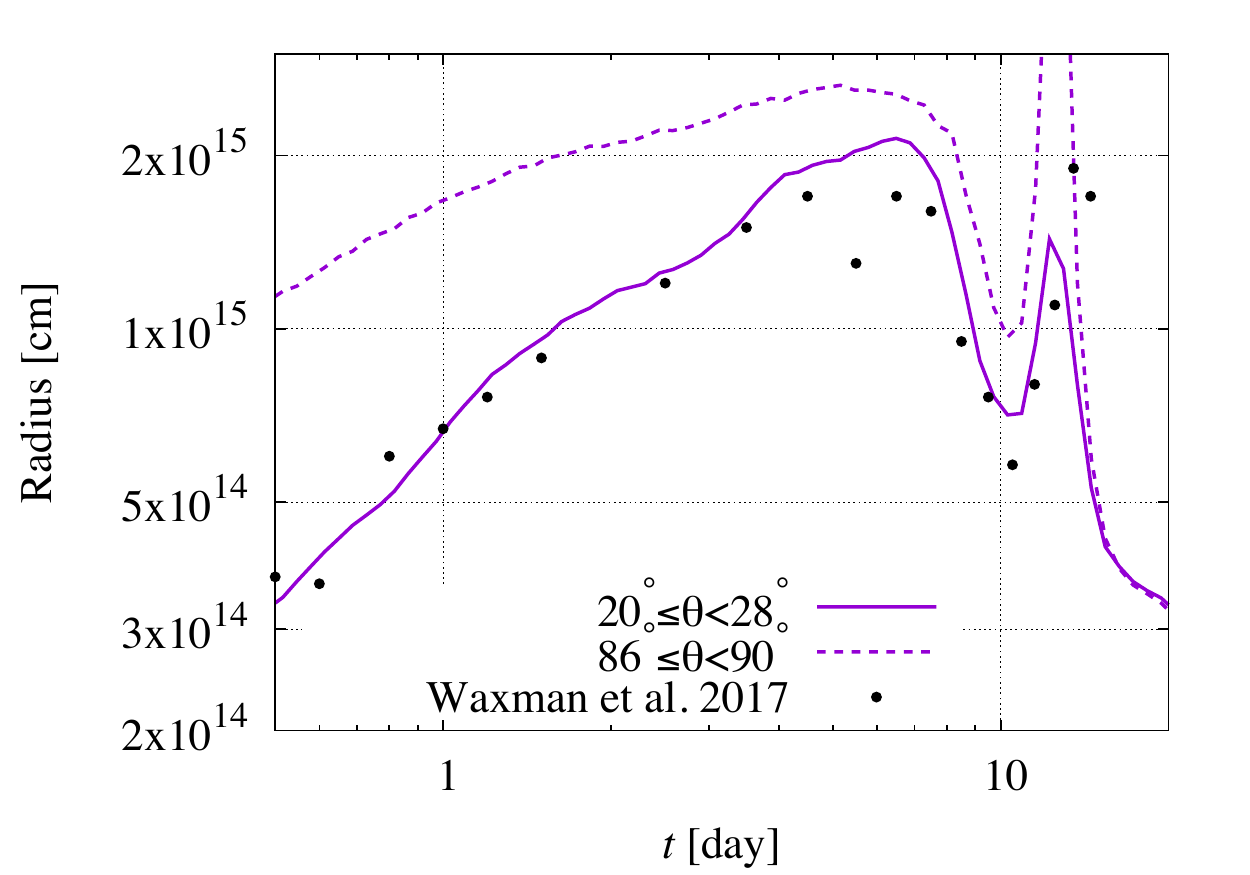}
 	 \includegraphics[width=.45\linewidth]{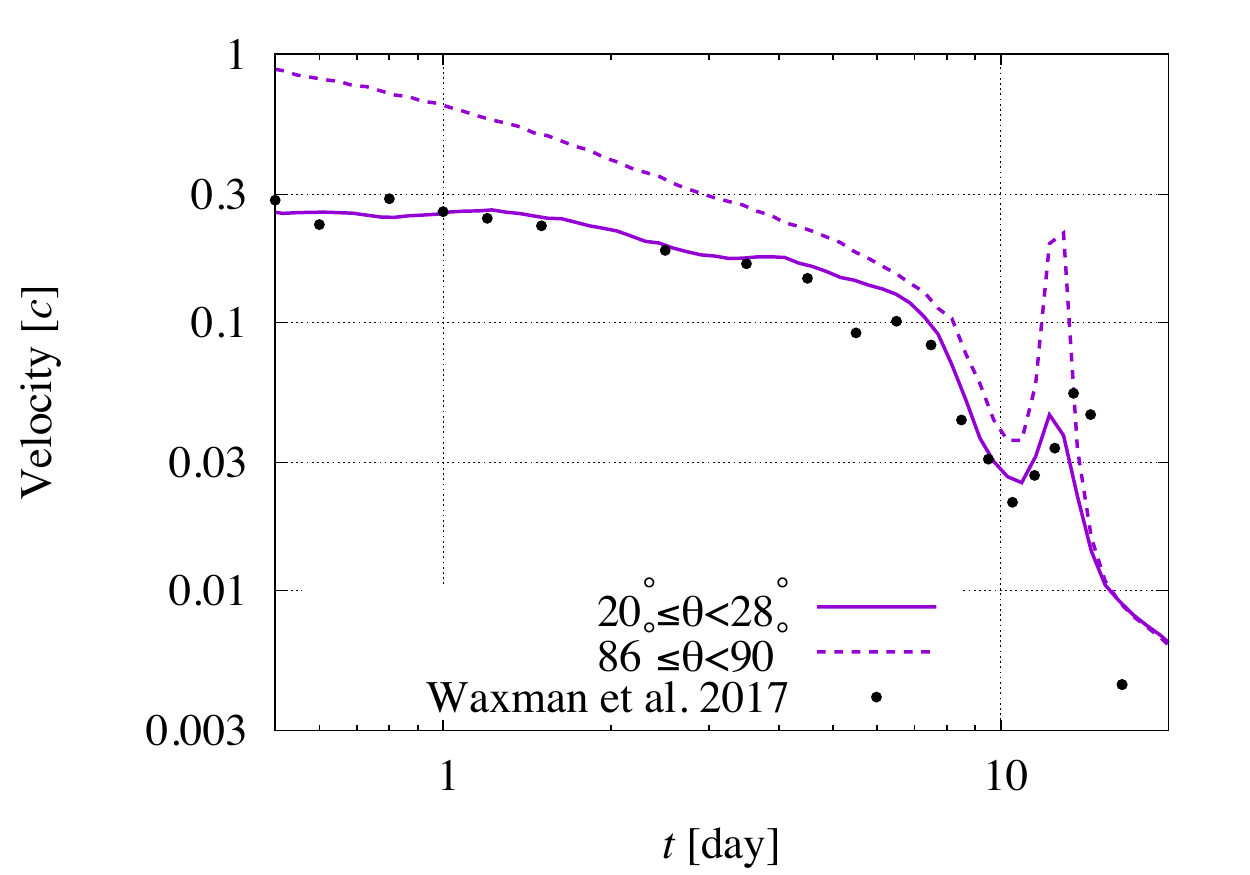}
 	 \caption{Bolometric luminosity ($L_{\rm bol}$; Top left), effective temperature ($T_{\rm eff}$; Top right), photospheric radius ($r_{\rm ph}$; Bottom left), and photospheric velocity ($v_{\rm ph}$; Bottom right) of the kilonova/macronova model of SSS17a. $L_{\rm bol}$ and $v_{\rm ph}$ are calculated by $L_{\rm bol}=4\pi r_{\rm ph}^2 \sigma T_{\rm eff}^4$ and $v_{\rm ph}=r_{\rm ph}/t$, respectively, using $T_{\rm eff}$ and $r_{\rm ph}$ obtained by the blackbody fit of the spectra. The solid and dashed curves denote the quantities calculated from the lightcurves observed from $20^\circ\le\theta\le28^\circ$ and $86^\circ\le\theta\le90^\circ$, respectively. The black points denote the data points of SSS17a taken from~\cite{Waxman:2017sqv}.}\label{fig:photo}
\end{figure*}

Figure~\ref{fig:spec} shows the time evolution of optical and NIR spectral energy distribution of the kilonova/macronova model observed from $20^\circ\le\theta\le28^\circ$. As shown in the observation of SSS17a~\cite[e.g.,][]{Waxman:2017sqv}, the spectra of our model agree approximately with blackbody spectra for $t\approx$1--7 days. Figure~\ref{fig:photo} shows the (isotropic) bolometric luminosity, $L_{\rm bol}$, effective temperature, $T_{\rm eff}$, photospheric radius, $r_{\rm ph}$, and photospheric velocity, $v_{\rm ph}$, of the kilonova/macronova model. $T_{\rm eff}$ and $r_{\rm ph}$ are first obtained by the blackbody fit of the spectra, and then, $L_{\rm bol}$ and $v_{\rm ph}$ are calculated by $L_{\rm bol}=4\pi r_{\rm ph}^2 \sigma T_{\rm eff}^4$ and $v_{\rm ph}=r_{\rm ph}/t$, respectively, where $\sigma$ is the Stefan-Boltzmann constant. We find that all these quantities calculated from the lightcurves observed from $20^\circ\le\theta\le28^\circ$ agree with the observation~\citep{Waxman:2017sqv}. In particular, $v_{\rm ph}\approx0.3\,c$ is realized for $t\le2$ days due to photons reprocessed in the dynamical ejecta. We here stress that the presence of the low-density dynamical ejecta in the polar region is the key to interpret the observed value of $v_{\rm ph}$. Indeed, we find that the value of $v_{\rm ph}$ cannot be as large as $0.25\,c$ for $\agt 1$ day if the low-density dynamical ejecta region in $\theta\le\pi/4$ is absent. 

Figure~\ref{fig:photo} also shows the photospheric quantities calculated from the lightcurves observed from the equatorial direction ($86^\circ\le\theta\le90^\circ$). The luminosity and temperature of the lightcurves are lower than those observed from $20^\circ\le\theta\le28^\circ$ by a factor of 3--4 and 2--3 at $\sim1$ day, respectively, and a larger radius and higher velocity are realized for the photosphere. These differences clearly reflect the density and velocity profiles of ejecta such that optically thick dynamical ejecta in the equatorial plane is present outside the post-merger ejecta.

\begin{figure}
 	 \includegraphics[width=.95\linewidth]{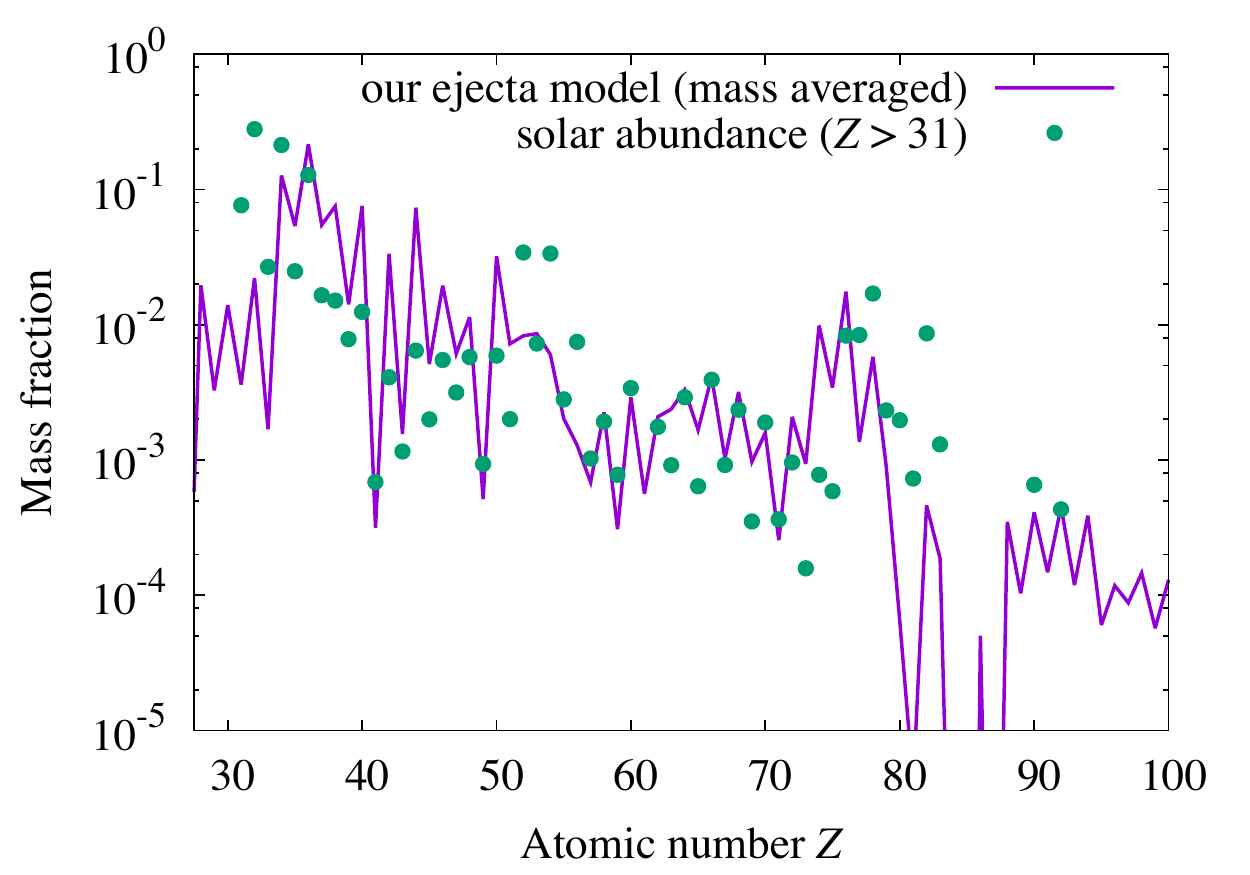}
 	 \caption{The mass-averaged element abundance of our model (blue lines) together with the solar abundance of r-process element~\citep[][green points]{Simmerer:2004jq}.}\label{fig:abun}
\end{figure}
NS mergers are also considered to be important synthesis sites of r-process nuclei in the universe~\citep{Lattimer:1974slx,Eichler:1989ve,Korobkin:2012uy,Wanajo:2014zka}. Figure~\ref{fig:abun} compares the elemental abundance in our model with the solar abundance. Though some abundance peaks are smaller than those of the solar abundance, broadly speaking, the mass-averaged element abundance of our model reproduces the trend of the solar abundance for a wide range of r-process elements, in particular, including the first ($Z$ = 34) abundance peaks. 

\section{Discussion \& Conclusion}\label{sec:sec4}
In this letter, we showed a result of an axisymmetric radiative transfer simulation for a kilonova/macronova with a setup indicated by numerical-relativity simulations. In particular, the interplay of multiple non-spherical ejecta components via photon transfer are consistently taken into account in the lightcurve prediction. 

We found that the optical and NIR lightcurves of SSS17a are reproduced naturally by the numerical-relativity-simulation-motivated model observed from $20^\circ\le\theta\le28^\circ$. In particular, we showed that the observed NIR lightcurves can be interpreted by the emission from the dynamical ejecta of which mass is consistent with the prediction of numerical relativity. The observed lightcurves are reproduced by a smaller mass of the post-merger ejecta than that estimated by previous studies~\citep[e.g.,][]{Kasliwal:2017ngb,Cowperthwaite:2017dyu,Kasen:2017sxr,Perego:2017wtu,Villar:2017wcc} because the effect of the photon diffusion preferentially to the polar direction is taken into account. The observed blue optical lightcurves as well as the photospheric velocity of $\approx 0.3\,c$ can be interpreted by the photon-reprocessing in the low-density dynamical ejecta, which locates in the polar region above the post-merger ejecta.

Our results indicate that there is no tension between the prediction of numerical-relativity simulations and the observation of SSS17a, and that the interplay of the multiple non-spherical ejecta components plays a key role for predicting kilonova/macronova lightcurves. Note that~\cite{Perego:2017wtu} showed a semi-analytical model for
kilonova/macronova by employing a similar setup with our model but did not discuss the high photospheric velocities. Note that our model requires $\sim 0.01 M_\odot$ as the mass of the dynamical ejecta. This is a fairly large value for the dynamical ejecta, which can be achieved only for the case that the NS radii are small~\citep[e.g.,][]{Hotokezaka:2012ze,Dietrich:2016hky}. Thus our analysis suggests that the NS radius would be small as $\alt 12$\,km.

We found that photons from the post-merger ejecta are absorbed and entirely reprocessed by the dynamical ejecta in particular if the binary is observed from the equatorial direction. However, this viewing angle dependence would be minor for the case that the total mass of the binary is smaller than GW170817. For such a case, the mass of the dynamical ejecta would be much smaller ($\sim 10^{-3} M_\odot$ or less)~\citep{Foucart:2015gaa}, and thus, suppression of the blue optical emission would be weaker. Furthermore, a long-lived remnant NS is likely to be formed after the merger, and the lightcurves could be significantly modified by the heating up of the ejecta due to the EM radiation from the strongly-magnetized and rapidly rotating remnant NS~\citep[e.g.,][]{Metzger:2013cha}.

While our kilonova/macronova model of SSS17a agrees approximately with the observation, some deviation from the data points, for example $\agt2$ mag differences in the {\it ugri} and {\it zJHK}-band for $t\ge3$ days and $t\ge11$ days, respectively, is also found. This may be due to the simplification of the $Y_e$ distribution in our model in which we neglect its local dependence found in the simulations~\citep[e.g.,][]{Sekiguchi:2016bjd,Bovard:2017mvn,Metzger:2014ila,Fujibayashi:2017puw}. We suspect that the deviation of the mass-averaged abundance pattern from the observation found in Figure~\ref{fig:abun} might be due to the same reason. The incompleteness of the line list for the opacity estimation is also an issue. For example, we suspect that the large deviation found in the model lightcurves in {\it H}-band may be due to the simplification that the same bound-bound transition properties are used for the elements with the same open shell. Thus, employing a detailed ejecta profile based on numerical-relativity simulations and more realistic opacity table are needed to reproduce the observation more accurately. 
\begin{acknowledgments}
We thank S. Fujibayashi, K. Hotokezaka, and E. Waxman for valuable discussions. Numerical computation was performed on Cray XC30 at cfca of National Astronomical Observatory of Japan and on Hydra at Max Planck Computing and Data Facility. This work was supported by Grant-in-Aid for Scientific Research (JP16H02183, JP16H06342, JP17H01131, JP15K05077, JP17K05447, JP17H06361, JP15H02075, JP17H06363 ) of JSPS and by a post-K computer project (Priority issue No. 9) of Japanese MEXT. K. K. was supported by JSPS overseas research fellowships.
\end{acknowledgments}

\bibliographystyle{apj}

\end{document}